\documentclass[12pt]{article}
\usepackage{graphicx}
\usepackage{hyperref}
\usepackage{amsmath}
\usepackage{amscd,amssymb}
\usepackage{xcolor}

\newcommand{\fz}{\mathfrak{z}}

\newcommand{\id}{{1\!\!1}}

\def\cR{{\mathcal R}}

\def\cO{{\mathcal O}}

\def\mB{{\mathfrak B}}

\def\mg{{\mathfrak g}}

\def\mg{{\mathfrak g}}

\newtheorem{remark}{Remark}[section]

\newcommand{\beq}{\begin{eqnarray}}
\newcommand{\eeq}{\end{eqnarray}}
\numberwithin{equation}{section}

\begin{document}


\begin{center}
{\large\bf Vector Generation Functions, q-Spectral Functions of Hyperbolic Geometry and Vertex Operators 
for Quantum Affine Algebras}
\end{center}

\vspace{0.1in}

\begin{center}
{\large
A. A. Bytsenko $^{(a)}$
\footnote{E-mail: aabyts@gmail.com},
M. Chaichian $^{(b)}$
\footnote{E-mail: masud.chaichian@helsinki.fi}
and R. Luna $^{(a)}$
\footnote{E-mail: rodrigo.mluna@outlook.com}}

\vspace{0.5cm}
$^{(a)}$
{\it
Departamento de F\'{\i}sica, Universidade Estadual de
Londrina\\ Caixa Postal 6001,
Londrina-Paran\'a, Brazil}

\vspace{0.2cm}
$^{(b)}$
{\it
Department of Physics, University of Helsinki\\
P.O. Box 64, FI-00014 Helsinki, Finland}

\end{center}

\begin{abstract}
We investigate the concept of $q$-replicated arguments in symmetric functions 
with its connection to spectral functions of hyperbolic geometry.
This construction suffices for vector generation functions in the form of $q$-series, 
and string theory. We hope that the mathematical side of the construction can be 
enriched by ideas coming from physics.

\end{abstract}

\vspace{0.1in}

\begin{flushleft}
PACS \, 02.10. Kn, 02.20.Uw, 04.62.+v

\vspace{0.3in}
July 2017
\end{flushleft}

\newpage

\tableofcontents

\vspace{0.1in}

\section{Introduction}

In this paper we discuss the multipartite ({\it vector}) generation functions, vertex operators, 
and how it can be applied to derive some explicit results concerning the bosonic
strings, symmetric functions, spectral functions of hyperbolic geometry, and
vertex operator traces.

{\bf The organization of the paper and our key results.}
In Sect. 2 we begin with a brief review of the multipartite generation functions 
and derive some explicit results using the Bell polynomial technique. 

We then turn to the polynomial ring $\Lambda(X)$ and its algebraic properties (Sect. 3)
and proceed to apply restricted specializations and $q$-series. 

$2N$-piecewise string we consider in Sect. 4.1. A piecewise uniform bosonic string which 
consists of $2N$ parts of equal length, of alternating type I and type II material, is relativistic 
in the sense that the velocity of sound everywhere equals the velocity of light. 
The present section is a continuation of two earlier papers, one dealing with the Casimir energy 
of a $2N$-piece string \cite{Brevik90}, and 
another dealing with the thermodynamic properties of a string divided into two (unequal) parts 
\cite{Brevik98}. 

Finally in Sects. 4.2 and 4.3 we turn to symmetric functions with replicated variables,
and vertex operator traces with it connection to the spectral functions of hyperbolic geometry. 
There have been interesting developments in physics, in string theory, and related subjects.
During the last years or so, the mathematical side has been greatly enriched by ideas from physics.

\section{Multipartite generating functions}
\label{Mult}

For any ordered $m$-tuple or {\it multipartite} numbers of nonnegative integers (not all zeros),
$(k_1, k_2, \ldots ,k_m)={\overrightarrow{k}}$ let us consider the
(multi)partitions, i.e. distinct representations of $(k_1, k_2, \ldots ,k_m)$ as sums
of multipartite numbers. Let
${\mathcal C}_-^{(z,m)}({\overrightarrow{k}}) = {\mathcal C}_-^{m}(z;k_1, k_2 , \cdots , k_m)$
be the number of such multipartitions; in addition introduce the symbol
${\mathcal C}_+^{(z,m)} ({\overrightarrow{k}})= {\mathcal C}_+^{(m)}(z;k_1, k_2 , \cdots , k_m)$.
Their generating functions can be defined as \cite{Andrews}
\begin{eqnarray}
{\mathcal F}({z;X}) & \stackrel{{\rm def}}{=} & \prod_{{\overrightarrow{k}}\geq 0} \left( 1- z x_1^{k_1}x_2^{k_2}
\cdots x_m^{k_m}\right)^{-1}
= \sum_{{\overrightarrow{k}}\geq 0}{\mathcal C}_-^{(z,m)}({\overrightarrow{k}})
x_1^{k_1}x_2^{k_2}\cdots x_r^{k_m}\,,
\label{PF1}
\\
{\mathcal G }(z;X) &\stackrel{{\rm def}}{=}& \prod_{{\overrightarrow{k}}\geq 0} \left( 1 + z x_1^{k_1}x_2^{k_2}
\cdots x_m^{k_m}\right)
= \sum_{{\overrightarrow{k}}\geq 0}{\mathcal C}_+^{(z,m)}({\overrightarrow{k}})
x_1^{k_1}x_2^{k_2}\cdots x_m^{k_m}\,.
\label{PF2}
\end{eqnarray}
Then
\begin{eqnarray}
{\rm log}\, {\mathcal F}(z;X) & = & - \sum_{{\overrightarrow{k}}\geq 0} {\rm log}
\left(1- z x_1^{k_1}x_2^{k_2}\cdots x_r^{k_m}\right)
=  \sum_{{\overrightarrow{k}}\geq 0} \sum_{n=1}^\infty \frac{z^n}{n}
x_1^{nk_1}x_2^{nk_2}\cdots x_m^{nk_m}
\nonumber \\
& = &
\sum_{n =1}^\infty \frac{z^n}{n}
(1-x_1^n)^{-1}(1-x_2^n)^{-1} \cdots (1-x_m^n)^{-1}
\nonumber \\
& = &
\sum_{n=1}^\infty \frac{z^n}{n}
\prod_{j= 1}^m (1-x_j^n)^{-1},
\\
{\rm log}\,{\mathcal G}(-z;X) & = & {\rm log}\,{\mathcal F}(z;X)\,.
\end{eqnarray}
Let $\beta_m(n) := \prod_{j=1}^m(1-x_j^n)^{-1}$, finally
\begin{eqnarray}
\!\!\!\!\!\!\!\!\!\!\!\!\!
{\mathcal F}(z;X) & = &
\sum_{{\overrightarrow{k}}\geq 0}{\mathcal C}_-^{(z,m)}({\overrightarrow{k}})
x_1^{k_1}x_2^{k_2}\cdots x_m^{k_m}
= \exp\left( \sum_{n=1}^\infty \frac{z^n}{n}
\beta_m(n)\right),
\\
\!\!\!\!\!\!\!\!\!\!\!\!\!
{\mathcal G }(z;X) & = &
\sum_{{\overrightarrow{k}}\geq 0}{\mathcal C}_+^{(z,m)}({\overrightarrow{k}})
x_1^{k_1}x_2^{k_2}\cdots x_m^{k_m}
= \exp\left( \sum_{n=1}^\infty \frac{(-z)^n}{n}
\beta_m(n)\right).
\end{eqnarray}

\subsection{The Bell polynomials}

The Bell polynomials, first extensively studied by E. T. Bell (see for example \cite{Bell}), arise in
the task of taking the $n$-th derivative of a composite function. Namely,
one can find a formula for the n-th derivative of h(t)= f(g(t)). 
If we denote $d^nh/dt^n= h_n, d^nf/dg^n=f_n,d^ng/dt^n = g_n$, 
then we see that $h_1 = f_1, h_2 = f_1g_2 +f_2g_1^2, h_3 = f_1g_3 + 3f_2g_2g_1
+f_3g_1^3, \cdots$. By mathematical induction we find that
$h_n  =  f_1\alpha_{n1}(g_1,\ldots,g_n) + f_2\alpha_{n2}(g_1,\ldots,g_n) 
+\cdots + f_n\alpha_{nn}(g_1,\ldots,g_n)$,
where $\alpha_{nj}(g_1,\ldots,g_n)$ is a homogeneous polynomial of degree $j$ in $g_1,\ldots,g_n$.

As a result, the study of $h_n$ may be reduced to the study of the {\it Bell polynomials}:
$Y_n(g_1, g_2, \ldots, g_n) =\alpha_{n1}(g_1, \ldots, g_n) + \alpha_{n2}(g_1, \ldots, g_n)
 + \cdots + \alpha_{nn}(g_1, \ldots, g_n)$.
Note that $Y_n$ is a polynomial in $n$ variables and the fact that $g_j$ was originally
an $j$-th derivative is not necessary in the consideration.

Recurrence relations for the Bell polynomial $Y_{n}(g_1, g_2, \ldots , g_{n})$
and generating function ${\mathcal B}(z)$ have the forms \cite{Andrews}:
\begin{equation}
Y_{n+1}(g_1, g_2, \ldots , g_{n+1}) = \sum_{k=0}^n  {\begin{pmatrix} n\cr k\end{pmatrix}}
Y_{n-k}(g_1, g_2, \ldots , g_{n-k})g_{k+1},
\label{B2}
\end{equation}
\begin{equation}
{\mathcal B}(z) = \sum_{n=0}^\infty Y_n z^n/n! \Longrightarrow
{\rm log}\,{\mathcal B}(z)= \sum_{n=1}^\infty g_n z^n/n!. 
\end{equation}

From the last formula one can obtain the following explicit
formula for the Bell polynomials (it is known as Faa di Bruno's formula)
\begin{equation}
Y_{n}(g_1, g_2, \ldots , g_{n}) = \sum_{{\bf k}\,\vdash\, n}\frac{n!}{k_1!\cdots k_n!}
\prod_{j=1}^n\left(\frac{g_j}{j!}\right)^{k_j}\!\!.
\end{equation}

If we let
\begin{eqnarray}
\!\!\!\!\!\!&& {\mathcal F}(z;X) := 1 + \sum_{j=1}^\infty {\mathcal P}_j(x_1,x_2, \ldots, x_m)z^j,
\,\,\,\,\,\,\,\,\,\,
{\mathcal P}_j  =  1+ \sum_{{\overrightarrow{k}}> 0}P({\overrightarrow{k}}; j)
x_1^{k_1}\cdots x_m^{k_m},
\label{Fu}
\\
\!\!\!\!\!\!&& {\mathcal G}(z;X)  :=  1 + \sum_{j=1}^\infty {\mathcal Q}_j(x_1,x_2, \ldots, x_r)z^j,
\,\,\,\,\,\,\,\,\,\,
{\mathcal Q}_j  =  1+ \sum_{{\overrightarrow{k}}> 0}Q({\overrightarrow{k}}; j)
x_1^{k_1}\cdots x_m^{k_m},
\label{Gu}
\end{eqnarray}
then the following result holds (see for detail \cite{Andrews}):
\begin{eqnarray}
{\mathcal P}_j  & = & \frac{1}{j!}Y_j \left( 0!\beta_m(1),\,\, 1!\beta_m(2)\,\,,
\ldots , \,\,(j-1)!\beta_m(j)\right),
\label{P1}
\\
{\mathcal Q}_j  & = & \frac{1}{(-1)^jj!}Y_j \left( -0!\beta_m(1),\,\,
-1!\beta_m(2)\,\,, \ldots , \,\,-(j-1)!\beta_m(j)\right).
\label{Q1}
\end{eqnarray}

\subsection{Restricted specializations and q-series}

Setting $X= (x_1, x_2, \ldots,x_r, 0, 0, \ldots)= (q, q^2, \ldots, q^r, 0, 0, \ldots)$ 
for finite additive manner, as a result we get
\begin{eqnarray}
{\mathcal F}(z;X) & = &
\prod_{{\overrightarrow{k}}\geq 0} \left( 1- z q^{k_1+k_2+\cdots + k_r}\right)^{-1}
= \exp\left( - \sum_{m=1}^\infty \frac{z^m}{m}\prod_{\ell = 1}^r(1-q^{\ell m})^{-1}\right),
\\
{\mathcal G}(z;X) & = &
\prod_{{\overrightarrow{k}}\geq 0} \left( 1+ z q^{k_1+k_2+\cdots + k_r}\right)
= \exp\left( - \sum_{m =1}^\infty \frac{(-z)^m}{m}\prod_{\ell = 1}^r(1-q^{\ell m})^{-1}\right).
\end{eqnarray}

{\bf Spectral functions of hyperbolic geometry.}
Let us begin by explaining the general lore for the characteristic classes and $\mg$-structure
on compact groups.

\begin{remark}
Suppose $\mathfrak g$ is the Lie algebra of a Lie group $G$. Let us
consider the pair $({\Gamma}, G)$ of Lie groups, where $\Gamma$ is a closed
subgroup of $G$ with normalizer subgroup $N_{\Gamma}\subset G$. Then the
pair $({\Gamma}, G)$ with the discrete quotient group $N_{\Gamma}/{ \Gamma}$ corresponds to the
inclusion ${\mathfrak g}\hookrightarrow W_n$, where $W_n$ is the Lie algebra of formal vector
fields in $n = {\rm dim} \,G/{\Gamma}$ variables, while the homogeneous space $G/{\Gamma}$
possesses a canonical $\mathfrak g$-{structure} $\omega$. Combining this $\mg$-structure with the
inclusion $\mg\hookrightarrow W_n$, one obtains a $W_n$-structure on
the quotient space $G/\Gamma$ for any discrete subgroup $\Gamma$ of the Lie group $G$; this is
precisely the $W_n$-structure which corresponds to the $\Gamma$-equivariant foliation
of $G$ by left cosets of $\Gamma$~{\rm \cite{Fuks}}.
The homomorphism
\begin{equation}
{{\rm char}_\omega :} \, \Gamma^\sharp (W_n)\rightarrow \Gamma^\sharp (G/\Gamma,
{\mathbb R})
\end{equation}
associated with characteristic classes of $W_n$-structures
decomposes into the composition of two homomorphisms
\begin{equation}
\Gamma^\sharp (W_n)\rightarrow \Gamma^\sharp ({\mathfrak g})
\,\,\,\,\, {\rm and}\,\,\,\,\,
\Gamma^\sharp ({\mathfrak g}) \rightarrow \Gamma^\sharp (G/\Gamma, {\mathbb R}).
\end{equation}
The first homomorphism is independent of $\Gamma$
and is induced by the inclusion ${\mathfrak g}\hookrightarrow W_n$,
while the second homomorphism is independent of $\Gamma$ and corresponds to the
canonical homomorphism which determines
the characteristic classes of the canonical $\mathfrak g$-structure $\omega$ on $G/\Gamma$.

If the group $G$ is semi-simple, then the Lie algebra $\mathfrak g$ is
unitary and $G$ contains a discrete subgroup $\Gamma$ for which
$G/\Gamma$ is compact; for appropriate choice of $\Gamma$ the kernel of the homomorphism
$\Gamma^\sharp (W_n)\rightarrow \Gamma^\sharp (G/\Gamma, {\mathbb R})$
coincides with the kernel of the homomorphism
$\Gamma^\sharp (W_n)\rightarrow \Gamma^\sharp ({\mathfrak g})$.
\end{remark}

In our applications we shall  consider a {\it compact hyperbolic} three-manifold $G/\Gamma$
with $G = SL(2, {\mathbb C})$.
By combining the characteristic class representatives of field theory, elliptic
genera with the homomorphism
$
{{\rm char}_\omega}
$,
we can compute quantum partition functions in terms of the
spectral functions of hyperbolic three-geometry {\rm\cite{BB,BCST}}.

Let us introduce next the Ruelle spectral function ${\mathcal R}(s)$ associated with
hyperbolic three-geometry \cite{BB,BCST}. The function ${\mathcal R}(s)$ is an
alternating product of more complicate factors, each of which is so-called Patterson-Selberg
zeta-functions $Z_{\Gamma^\gamma}$;
functions ${\mathcal R}(s)$ can be continued meromorphically to
the entire complex plane $\mathbb C$.
\begin{eqnarray}
\prod_{n=\ell}^{\infty}(1- q^{an+\varepsilon})
& = & \prod_{p=0, 1}Z_{\Gamma^\gamma}(\underbrace{(a\ell+\varepsilon)(1-i\varrho(\vartheta))
+ 1 -a}_s + a(1 + i\varrho(\vartheta)p)^{(-1)^p}
\nonumber \\
& = &
\cR(s = (a\ell + \varepsilon)(1-i\varrho(\vartheta)) + 1-a),
\label{R}
\\
\prod_{n=\ell}^{\infty}(1+ q^{an+\varepsilon})
& = &
\prod_{p=0, 1}Z_{\Gamma^\gamma}(\underbrace{(a\ell+\varepsilon)(1-i\varrho(\vartheta)) + 1-a +
i\sigma(\vartheta)}_s
+ a(1+ i\varrho(\vartheta)p)^{(-1)^p}
\nonumber \\
& = &
\cR(s = (a\ell + \varepsilon)(1-i\varrho(\vartheta)) + 1-a + i\sigma(\vartheta))\,,
\label{R2}
\end{eqnarray}
where $q\equiv e^{2\pi i\vartheta}$, $\varrho(\vartheta) =
{\rm Re}\,\vartheta/{\rm Im}\,\vartheta$,
$\sigma(\vartheta) = (2\,{\rm Im}\,\vartheta)^{-1}$,
$a$ is a real number, $\varepsilon, b\in {\mathbb C}$, $\ell \in {\mathbb Z}_+$.

Obviously, $\prod_{\ell = 1}^r(1-q^{\ell m})^{-1} \equiv
\prod_{\ell = 1}^\infty(1-q^{\ell m})^{-1}\prod_{\ell = r+1}^\infty(1-q^{\ell m})$ and
\begin{eqnarray}
{\mathcal F}(z;X) & = &
\prod_{{\overrightarrow{k}}\geq 0} \left( 1-  zq^{k_1+k_2+\cdots + k_r}\right)^{-1}
\nonumber \\
& = &
\exp\left( - \sum_{m=1}^\infty \frac{z^m}{m}\cdot
\frac{\cR(s= -im\varrho(\vartheta)(r+1)+mr+1)}
{\cR(s = -im\varrho(\vartheta)+ 1)}\right),
\label{F1}
\\
{\mathcal G}(z;X) & = &
\prod_{{\overrightarrow{k}}\geq 0} \left( 1+ z q^{k_1+k_2+\cdots + k_r}\right)
\nonumber \\
& = &
\exp\left( - \sum_{m =1}^\infty \frac{(-z)^m}{m}\cdot
\frac{\cR(s= -im\varrho(\vartheta)(r+1)+mr+1)}
{\cR(s = -im\varrho(\vartheta)+ 1)}\right)\,.
\label{G1}
\end{eqnarray}

{\bf Hierarchy.}
Setting $\cO q^{k_0+k_{1}+\ldots + k_{r}}= \cO_{{\overrightarrow{k}}}q^{k_{0}}$
with $\cO_{{\overrightarrow{k}}}=
\cO q^{k_{1}+\ldots +k_{r}}$\, (${\overrightarrow{k}} =
\left( k_{1},\ldots ,k_{r}\right))$ we get
\begin{equation}
Z_{2}\left( \cO_{{\overrightarrow{k}}},q\right) =\prod_{k_{0}=0}^{\infty}
\left[1-\cO_{\overrightarrow{k}}q^{k_{0}}\right]^{-1} =
[(1-\cO_{\overrightarrow{k}}){\mathcal R}(s= (k_{1}+\ldots + k_{r})(1-i\varrho(\tau)))]^{-1}\,.
\label{CFT_2}
\end{equation}
Therefore the infinite products
$\prod_{k_r=0}^\infty\prod_{k_{r-1}=0}^\infty\cdots\prod_{k_1=0}^\infty\prod_{k_0=0}^\infty
(1-q^{k_0+k_1+\cdots+k_r})^{-1}$ can be factorized as $
\prod_{{\overrightarrow{k}}\geq {\overrightarrow{0}}} Z_{2}\left(
\cO_{{\overrightarrow{k}}},q\right). $ We can treat this
factorization as a product of $r$ copies, each of them is
$Z_{2}\left( \cO_{{\overrightarrow{k}}},q\right)$ and corresponds
to a free two-dimensional conformal field theory (see \cite{BB}
for similar results).

\section{Symmetric functions for quantum affine algebras}
\label{Symmetric}

{\bf The polynomial ring $\Lambda(X)$.}
Let ${\mathbb Z}[x_1,\ldots,x_n]$ be the polynomial ring, or the
ring of formal power series, in $n$ commuting variables $x_1,\ldots,x_n$. The symmetric group
$S_n$ acting on  $n$ letters of this ring by permuting the variables.
For $\pi\in S_n$ and $f \in {\mathbb Z}[x_1,\ldots,x_n]$ we have
$
\pi f(x_1,\ldots,x_n) = f(x_{\pi(1)},\ldots,x_{\pi(n)}).
$
We are interested in the subring of functions invariant under this
action, $\pi f = f$, that is to say the ring of symmetric polynomials in
$n$ variables:
$
\Lambda(x_1,\ldots,x_n) = {\mathbb Z}[x_1,\ldots,x_n]^{S_n}.
$
This ring may be graded by the degree of the polynomials, so that
$
\Lambda(X)=\oplus_n\ \Lambda^{(n)}(X)
$,
where $\Lambda^{(n)}(X)$ consists of homogenous symmetric polynomials  in $x_1,\ldots,x_n$ of 
total degree $n$.

In order to work with an arbitrary number of variables, following
Macdonald~\cite{Macdonald}, we define the ring of symmetric functions
$\Lambda = \lim_{n\rightarrow\infty}\Lambda(x_1,\ldots,x_n)$ in its stable limit  ($n\rightarrow
\infty$). There exist various bases of $\Lambda(X)$:

{\bf (i)} A $\mathbb Z$ bases for $\Lambda^{(n)}$ are provided by the
monomial symmetric functions $\{m_\lambda\}$, where $\lambda$ is any partitions of $n$.

{\bf (ii)} The other (integral and rational) bases for $\Lambda^{(n)}$ are provided by the 
partitions $\lambda$ of $n$. There are the complete,
elementary and power sum symmetric functions:
$h_{\lambda}=h_{\lambda_1}h_{\lambda_2}\cdots h_{\lambda_n}$,
$e_{\lambda}=e_{\lambda_1}e_{\lambda_2}\cdots e_{\lambda_n}$ and
$p_{\lambda}=p_{\lambda_1}p_{\lambda_2}\cdots p_{\lambda_n}$, where for $\forall n \in {\mathbb Z}_+$, 
\begin{equation}
h_n(X) = \sum_{i_1\leq i_2\cdots\leq i_n}x_{i_1}x_{i_2}\cdots x_{i_n},\,\,\,\,\,\,\,\,
e_n(X) = \sum_{i_1<i_2\cdots<i_n}x_{i_1}x_{i_2}\cdots x_{i_n},\,\,\,\,\,\,\,\,
p_n(X) = \sum_{i}x_i^n,
\end{equation}
with the convention $h_0 = e_0 = p_0 =1,\, h_{-n} = e_{-n} = p_{-n} = 0$.
Three of these bases are multiplicative,
with $h_{\lambda}=h_{\lambda_1}h_{\lambda_2}\cdots h_{\lambda_n}$,
$e_{\lambda}=e_{\lambda_1}e_{\lambda_2}\cdots e_{\lambda_n}$ and
$p_{\lambda}=p_{\lambda_1}p_{\lambda_2}\cdots p_{\lambda_n}$.
The relationships between the various bases we just mention at this stage by
the transitions
\begin{equation}
p_\rho(X) = \sum_{\lambda\,\vdash n}\chi_\rho^\lambda s_\lambda(X)
\,\,\,\,\,\,\, {\rm and} \,\,\,\,\,\,\,
s_\lambda(X) = \sum_{\rho\,\vdash n}\ \fz_\rho^{-1}\ \chi^\lambda_\rho\ p_\rho(X)\,.
\label{Eq-p-s}
\end{equation}
For each partition $\lambda$, the Schur function is defined by
\begin{equation}
s_\lambda(X)\equiv s_\lambda(x_1,x_2, \ldots, x_n) = \frac{\sum_{\sigma\in S_n}{\rm sgn}
(\sigma)X^{\sigma(\lambda+\delta)}}{\prod_{i<j}(x_i-x_j)}\,,
\end{equation}
where $\delta = (n-1,n-2,\ldots,1,0)$. In fact both $h_n$ and $e_n$ are special Schur functions,
$h_n = s_{(n)},\, e_n = s_{(1^n)}$, and their generating functions are expressed in terms of 
the power-sum $p_n$:
\begin{equation}
\sum_{n\geq 0}h_nz^n = \exp (\sum_{n=1}^\infty(p_n/n)z^n), \,\,\,\,\,\,\,\,\,\,
\sum_{n\geq 0}e_nz^n = \exp (-\sum_{n=1}^\infty(p_n/n)(-z)^n)\,.
\end{equation}
The Jacobi-Trudi formula \cite{Macdonald} express the Schur functions in terms of $h_n$ or
$e_n$: $s_\lambda = {\rm det}(h_{\lambda_i-i+j}) = {\rm det}(e_{\lambda^\prime-i+j})$,
where $\lambda^\prime$ is the conjugate of $\lambda$. 
An involution $\omega: \Lambda\rightarrow \Lambda$ can be defined by $\omega(p_n) = (-1)^{n-1}p_n$. 
Then it follows that $\omega(h_n) = e_n$. Also we have $\omega(s_\lambda) = s_{\lambda^\prime}$.
$\chi^\lambda_\rho$ is the character of the irreducible representation
of the symmetric groups $S_n$ specified by $\lambda$ in the conjugacy class
specified by $\rho$. These characters satisfy the orthogonality conditions
\begin{eqnarray}
\sum_{\rho\,\vdash n}\ \fz_\rho^{-1}\ \chi^\lambda_\rho\ \chi^\mu_\rho = \delta_{\lambda,\mu}
\,\,\,\,\,\,\, {\rm and} \,\,\,\,\,\,\,
\sum_{\lambda\,\vdash n}\ \fz_\rho^{-1} \chi^\lambda_\rho \chi^\lambda_\sigma\ =
\delta_{\rho,\sigma}\,.
\label{Eq-chi-orth}
\end{eqnarray}
The significance of the Schur function bases lies in the fact that with respect
to the usual Schur-Hall scalar product
$\langle \cdot \,|\, \cdot \rangle_{\Lambda(X)}$ on $\Lambda(X)$ we have
\begin{eqnarray}
\langle s_\mu(X) \,|\, s_\nu(X) \rangle_{\Lambda(X)}
= \delta_{\mu,\nu}\,\,\,\,\,\, {\rm and}\,\,\,\,\, {\rm therefore}
\,\,\,\,\,
\langle p_\rho(X) \,|\, p_\sigma(X) \rangle_{\Lambda(X)}
= \fz_\rho \delta_{\rho,\sigma}\,,
\label{Eq-scalar-prod-p}
\end{eqnarray}
where $\fz_\lambda = \prod_i i^{m_i}m_i!$ for $\lambda = (1^{m_1}, 2^{m_2}, \cdots)$.

The ring, $\Lambda(X)$, of symmetric functions over $X$ has a Hopf algebra
structure, and two further algebraic and two coalgebraic operations. For
notation and basic properties we refer the reader to 
\cite{Fauser04,Fauser06} and references therein.

\section{Symmetric functions of a replicated argument}
\label{replicated}

\subsection{Example: 2N-piecewise string}

In this section we consider the bosonic {\it composite} string of length $L$ in $D$-dimensional spacetime,
which assumed to be uniform and consists of two or more uniform pieces. Such a model was introduced in 
1990 \cite{Brevik90}. Composite string was assumed to be divided into two pieces, of length $L_I$ and $L_{II}$, 
and it was relativistic in the sense that the velocity of sound was everywhere required to be equal 
to the velocity of light. Various aspects of the relativistic piecewise uniform string model 
were studied in \cite{Brevik98}. One may note, for instance, the paper of Lu and Huang \cite{lu} in which 
the model finds application in relation to the Green-Schwarz superstring.

The present paper focuses attention on the $2N$-piece string, made up of $2N$ parts of equal 
length, of alternating type I and type II material. 
The string of a total lenght $L$ is relativistic, the velocity of sound 
is everywhere equal to the velocity of light $v_s=\sqrt{T_I/\rho_I}=\sqrt{T_{II}/\rho_{II}}=c$,
where $T_I, T_{II}$ are the tensions and $\rho_I, \rho_{II}$ the mass densities in the two pieces. 

Our interest is the transverse oscillations $\psi=\psi(\sigma,\tau)$ of the string, where 
$\sigma$ denoting as usual the position coordinate and $\tau$ the time coordinate of the string. 
Thus in the two regions we have
\begin{equation}
\psi_I=\xi_I e^{i\omega (\sigma-\tau)}+\eta_I e^{-i\omega (\sigma+\tau)}, 
\,\,\,\,\,\,\,\,
\psi_{II}=\xi_{II} e^{i\omega (\sigma-\tau)}+\eta_{II} e^{-i\omega 
(\sigma+\tau)}\,,
\label{oscillators}
\end{equation}
where $\xi$ and $\eta$ are appropriate constants. The junction conditions say that $\psi$ itself 
as well as the transverse elastic force $T\partial \psi /\partial \sigma$ are continuous, i.e.
at each of the $2N$ junctions
\begin{equation}
\psi_I=\psi_{II},\,\,\,\,\,\,\,\,\,\,
T_I\partial \psi_I/\partial \sigma =T_{II}
\partial \psi_{II}/\partial \sigma\,.
\label{conditions}
\end{equation}

Define $x\stackrel{\rm def}{=} T_I/T_{II}$, and also the symbols $p_N$ and $\epsilon$ by
$p_N\stackrel{\rm def}{=} \omega L/N$,\,\,$\epsilon\stackrel{\rm def}{=} (1-x)/(1+x)$. 
\begin{itemize}
\item{}
The eigenfrequencies are determined from
$
{\rm det}\left({\bf M}_{2N}(x,p_N)- \id\right)=0.
$
Here it is convenient to scale the resultant matrix $ {\bf M}_{2N}$ as \cite{Brevik97}
\begin{eqnarray}
&& \!\!\!\!\!\!\!\!\!\!\!\!\!\!  {\mathbf M}_{2N}(x,p_N)  =  
\left( \frac{(1+x)^2}{4x}\right)^N{\mathbf m}_{2N}(\epsilon, p_N), \,\,\,\,\,
{\bf m}_{2N}(\epsilon, p_N)=\prod_{j=1}^{2N}{\bf m}^{(j)}(\epsilon, p_N),
\\
&& \!\!\!\!\!\!\!\!\!\!\!\!\!\! {\mathbf m}^{(j)}(\epsilon,p_N)  =  
\left( \begin{array}{ll}
                                   1, & \mp \epsilon e^{-ijp_N}\\
                                   \mp \epsilon e^{ijp_N}, & 1
                                     \end{array} \right)
\end{eqnarray}
for $j=1, 2,...(2N-1)$. The sign convention is to use +/- for even/odd $j$. 
\item{}
At the last junction, for $j=2N$, the component matrix has a particular form 
(given an extra prime for clarity):
$
{\mathbf m}'_{2N}(\epsilon,p_N)=\left( \begin{array}{ll}
                                   e^{-iN p_N}, & \epsilon e^{-i N p_N}\\
                                   \epsilon e^{iN p_N}, & e^{iN p_N}
                                    \end{array} \right) .
$ The recursion formula alluded to above can be stated:
\begin{equation}
{\mathbf m}_{2N}(\epsilon, p_N)= {\mathbf \varOmega}^N(\epsilon, p_N), \,\,\,\,\,\,
{\mathbf \varOmega}(\epsilon,p)=\left( \begin{array}{ll}
                                 a & b\\
                                 b^* & a^*
                                 \end{array} \right) ,
\end{equation}
with $a=e^{-ip}-\epsilon^2,\,b=\epsilon (e^{-ip}-1)$. The obvious way to proceed is now to 
calculate the eigenvalues of $\mathbf \varOmega$, and express the elements of 
$ {\mathbf M}_{2N}$ as powers of these. More details can be found in \cite{Brevik97}.
\item{}
Assume that $L=\pi$, in conformity with usual practice. Thus $p_N=\pi \omega/N$. We let 
$X^\mu(\sigma,\tau)$, with $\mu=0,1,2,.., (D-1)$, specify the coordinates on the world sheet. 
For each of the eigenvalue branches we can write $X^\mu$ on the form
\begin{eqnarray}
X^\mu & = & x^\mu+\frac{p^\mu \tau}{\pi {T_0}}+X_I^\mu,~~{\rm region ~I},
\label{I}
\\
X^\mu & = & x^\mu+\frac{p^\mu \tau}{\pi {T_0}}+X_{II}^\mu,~~{\rm region ~II},
\label{II}
\end{eqnarray}
where $x^\mu$ is the centre--of--mass position, $p^\mu$ is the total momentum 
of the string, and ${T_0}=\frac{1}{2}(T_I+T_{II})$
is the mean tension. Further, $X_I^\mu$ and $X_{II}^\mu$ can be decomposed 
into oscillator coordinates,
\begin{eqnarray}
X_I^\mu & = & \frac{i}{2}{\ell}_s \sum_{n \neq 0} \frac{1}{n}
\left( \alpha_{nI}e^{i\omega (\sigma-\tau)}+\tilde{\alpha}_{nI}
e^{-i\omega(\sigma+\tau)} \right),
\label{XI}
\\
X_{II}^\mu & = & \frac{i}{2}{\ell}_s\sum_{n\neq 0}\frac{1}{n}\left( 
\alpha_{nII}e^{i\omega (\sigma-\tau)}+\tilde{\alpha}_{nII}e^{-i\omega 
(\sigma+\tau)} \right).
\label{XII}
\end{eqnarray}
Here ${\ell}_s$ is the fundamental string length, unspecified so far, and 
$\alpha_n, \tilde{\alpha}_n$ are oscillator coordinates of the right- and 
left-moving waves, respectively. A characteristic property of the 
composite string is that the oscillator coordinates have to be specified 
for each of the various branches.  
\end{itemize}
A significant simplification can be obtained if, following Ref.~\cite{Brevik98}, we 
limit ourselves to the case of extreme string rations only. It is clear that 
the eigenvalue spectrum has to be invariant under the transformation $x \rightarrow 1/x$. 
It is sufficient, therefore, to consider the tension ratio interval $0<x \leq 1$ only. 
The case of extreme tensions corresponds to $x \rightarrow  0$. We consider only this case 
in the following.

Assume that the case $x \rightarrow 0$ corresponds to $T_I \rightarrow 0$. 
Also $\epsilon \rightarrow 1$ and  
$\lambda_- =0,~\lambda_+=\cos p_N-1$ (the case of extreme tensions \cite{Brevik98}). We obtain 
the remarkable simplification that all the eigenfrequency branches degenerate into one single 
branch determined by $\cos p_N=1$. Thus the eigenvalue spectrum becomes 
$\omega_n=2Nn, \,\, n=\pm 1, \pm 2, \pm 3,...$. The junction conditions (\ref{conditions}) 
permit all waves to propagate from region I to region II.

{\bf \bf Oscillator coordinates in $2N$-piecewise model.}
The case $x\rightarrow 0$ gives the equations $\xi_I+\eta_I = 
2\xi_{II}=2\eta_{II}$, which show that the right- and left- moving amplitudes $\xi_I$ and 
$\eta_I$ in region I can be chosen freely and the amplitudes 
$\xi_{II}, \eta_{II}$ in region II are fixed. This means, in oscillator language, 
that $\alpha_n^\mu$ and $\tilde{\alpha}_n^\mu$ can be chosen freely. 
Choosing the fundamental length equal to ${\ell}_s= (\pi T_I)^{-1/2}$,
we can write the expansion (\ref{XI}) and (\ref{XII}) in both regions 
as (subscript I and II on the $\alpha_n$'s omitted)
\begin{eqnarray}
X_I^\mu & = & \frac{i}{2\sqrt{\pi T_I}}\sum_{n\neq 0}\frac{1}{n}\left( 
\alpha_n^\mu e^{2iNn(\sigma-\tau)}+\tilde{
\alpha}_n^\mu e^{-2iNn(\sigma+\tau)} \right),
\label{XImu}
\\
X_{II}^\mu & = & \frac{i}{2\sqrt{\pi T_I}}\sum_{n\neq 0}\frac{1}{n}\gamma_n^\mu 
e^{-2iNn\tau}\cos(2Nn\sigma),
\label{XIImu}
\end{eqnarray}
where we have defined $\gamma_n^\mu$ as $\gamma_n^\mu=\alpha_n^\mu+\tilde{\alpha}_n^\mu,~~~n \neq 0$.

Recall that the string action is
\begin{equation}
S=-(1/2)\int d\tau d\sigma T(\sigma) \eta^{\alpha \beta}
\partial_\alpha X^\mu \partial_\beta X_\mu,
\end{equation}
where $\alpha, \beta =0,1$ and $T(\sigma)=T_I$ in region I,\, 
$T(\sigma)=T_{II}$ in region II. The momentum conjugate to $X^\mu$ is 
$P^\mu(\sigma)=T(\sigma)\dot{X}^\mu$, and the Hamiltonian is accordingly
\begin{equation}
H=\int_0^\pi \left( P_\mu(\sigma)\dot{X}^\mu- {\cal L} \right) d\sigma=
(1/2)
\int_0^\pi T(\sigma)(\dot{X}^2+X'^2)d\sigma,
\label{19}
\end{equation}
where ${\cal L}$ is the Lagrangian. Some care has to be taken for the string constraint 
equation. In the classical theory for the uniform string the constraint equation reads 
$T_{\alpha \beta}=0$,\, $T_{\alpha \beta}$ being the energy-momentum tensor. 
However the situation is here more complicated, since the junctions restrict the freedom, 
and one has to take the variations $\delta X^\mu$. Thus we have to replace the strong 
condition $T_{\alpha \beta}=0$ by a weaker condition. The most natural choice, which we 
will adopt, is to to impose that $H=0$ when applied to the physical states.

Let us introduce lightcone coordinates, $\sigma^-=\tau-\sigma$ and 
$\sigma^+=\tau+\sigma$. The derivatives conjugate to $\sigma^\mp$ are 
$\partial_\mp=\frac{1}{2}(\partial_\tau \mp \partial_\sigma)$. 
\begin{eqnarray}
&&{\rm Region \,\,I:}
\nonumber \\
&&\partial_-X^\mu  =  \frac{N}{\sqrt{\pi T_I}}\sum_{-\infty} ^\infty 
\alpha_n^\mu e^{2iNn(\sigma-\tau)}, \,\,\,\,
\partial_+X^\mu=\frac{N}{\sqrt{\pi T_I}}\sum_{-\infty}^\infty 
\tilde{\alpha}_ne^{-2iNn(\sigma+\tau)}.
\label{dXI}
\\
&&{\rm Region \,\,II:}
\nonumber \\ 
&&\partial_\mp X^\mu  = \frac{N}{2\sqrt{\pi T_I}}\sum_{-\infty}^\infty 
\gamma_n^\mu e^{\pm 2in(\sigma \mp\tau)},
\label{dXII}
\end{eqnarray}
where we have defined
$
\alpha_0^\mu=\tilde{\alpha}_0^\mu=\frac{p^\mu}{NT_{II}}
\sqrt{\frac{T_I}{\pi}}, \,
\gamma_0^\mu=2\alpha_0^\mu.
$
Inserting these expressions into the Hamiltonian
%
we obtain \cite{Brevik98}
\begin{equation}
H  =  \frac{1}{2}N^2\sum_{-\infty}^\infty (\alpha_{-n}\cdot \alpha_n 
+\tilde{\alpha}_{-n}\cdot \tilde{\alpha}_n) +
\frac{N^2}{4x}\sum_{-\infty}^\infty \gamma_{-n}\cdot \gamma_n .
\label{H1}
\end{equation}
The momentum conjugate to $X^\mu$ is at any position on the string equal 
to $T(\sigma)\dot{X}^\mu$. We accordingly require the commutation rules 
\begin{eqnarray}
T_I[\dot{X}^\mu (\sigma,\tau),X^\nu(\sigma',\tau)] & = & -i\delta (\sigma-\sigma')
\eta^{\mu\nu}, \,\,\,\,\, {\rm region} \,\, I,
\label{26}
\\
T_{II}[\dot{X}^\mu(\sigma,\tau), X^\nu(\sigma',\tau)] & = &-i\delta 
(\sigma-\sigma') \eta^{\mu\nu}, \,\,\,\,\, {\rm region} \,\, II,
\label{27}
\end{eqnarray}
$\eta^{\mu\nu}$ being the $D$-dimensional flat metric. The other commutators 
vanish. The quantities to be promoted to Fock state operators are 
$\alpha_{\mp n}$ and $\gamma_{\mp n}$. We insert the expansions for 
$X^\mu$ and $\dot{X}^\mu$ for regions I and II and 
make use of the effective relationship
$
\sum_{n=-\infty}^\infty e^{2iNn(\sigma -\sigma')}=2\sum_{n=-\infty}^\infty 
\cos 2Nn\sigma \cos 2Nn\sigma' 
\longrightarrow \frac{\pi}{N}\delta (\sigma -\sigma').
\label{28}
$
We then get 
$
[\alpha_n^\mu, \alpha_m^\nu]  =  n\delta_{n+m,0}\eta^{\mu\nu},
$ region I (with a similar relation for $\tilde{\alpha}_n$),
$ 
[\gamma_n^\mu, \gamma_m^\nu] = 4nx\,\delta_{n+m,0} \,\eta^{\mu\nu}
$, region II.

Introduce annihilation and creation operators by
\begin{equation}
\alpha_n^\mu=\sqrt{n}\, a_n^\mu,~~~\alpha_{-n}^\mu=\sqrt{n}\,
a_n^{\mu \dagger},\,\,\,\,\,
\gamma_n^\mu=\sqrt{4nx}\,c_n^\mu,~~~\gamma_{-n}^\mu=\sqrt{4nx}\,
c_n^{\mu \dagger},
\label{31}
\end{equation}
and find for $n \geq 1$ the standard form
\begin{equation}
[a_n^\mu, a_m^{\nu \dagger}]=\delta_{nm}\eta^{\mu\nu}, \,\,\,\,\,
[c_n^\mu,c_m^{\nu\dagger}]=\delta_{nm}\eta^{\mu\nu}.
\label{commutation}
\end{equation}
Note that infinite dimensional Heisenberg algebras play a central role 
in applying symmetric function techniques to various problems in
mathematical physics. Such algebra is generated by operators 
$\{\tilde{\alpha}_m^\mu, \alpha_n^\nu, \gamma_n^\nu\vert n,m\in {\mathbb Z}\}$
obeying the commutation relations of type (\ref {commutation}).These algebras can be
realized on the space of symmetric functions by the association
\begin{equation}
\alpha_{-n}= p_n(X), \,\,\,\,\alpha_n = n\frac{\partial}{\partial p_n(X)}
\,\,\,\,n>0,
\label{realization}
\end{equation}
with the central element $\alpha_0$ acting as a constant. An alternative 
basis to that consisting of monomials in the creation operators $\alpha_{-n}$,
which corresponds to the power sum basis $p_\lambda(X)$, is the basis 
consisting of all Schur functions $s_\lambda(X)$.The symmetric-function 
basis has proven convenient for carrying out calculations in bosonic Fock 
spaces, using the realization (\ref{realization}). In the case of two commuting copies 
$\{\alpha_{-n}, \tilde{\alpha}_{-m}\}$ of the Heisenberg algebra realized on the space 
$\Lambda(X)\times\Lambda(Y)$ for a state $\vert u \rangle =
\alpha_{-n_1}\cdots \alpha_{-n_p}\tilde{\alpha}_{-m_1}\cdots 
\tilde{\alpha}_{-m_r}\vert 0\rangle$ we obtain $\vert\vert u\vert\vert^2 =
\vert\vert\alpha_{-n_1}\cdots\alpha_{-n_p}\vert 0\rangle
\vert\vert^2\cdot\vert\vert\tilde{\alpha}_{-m_1}\cdots\tilde{\alpha}_{-m_r}
\vert0\rangle\vert\vert^2$. Therefore in the language of symmrtric functions 
this corresponds to using the inner product 
$\langle\cdots,\cdots\rangle_{\Lambda(X)\times\Lambda(Y)}$ on the space 
$\Lambda(X)\times\Lambda(Y)$.

\subsection{Replicated argument and spectral functions of hyperbolic geometry}
\label{skew functions}

The Schur-Hall scalar product may be used to define skew Schur functions
$s_{\lambda/\mu}$ through the identities
$
c^\lambda_{\mu,\nu}
= \langle s_\mu\, s_\nu \vert s_\lambda \rangle  = \langle s_\nu\,\vert s_\mu^\perp (s_\lambda)
\rangle = \langle s_\nu \vert s_{\lambda/\mu} \rangle\,,
$
so that
$
s_{\lambda/\mu} =  \sum_\nu\ c^\lambda_{\mu,\nu}\ s_\nu\,.
$
In what follows we shall make considerable use of several infinite series
of Schur functions. The most important of these are the mutually inverse
pair defined by
\begin{eqnarray}
 {\mathcal F}(t;X) & = & \prod_{i\geq 1} (1-t\,x_i)^{-1} = \sum_{m\geq 0} h_m(X)t^m\,
\label{Eq-M}\\
{\mathcal G}(t;X)& = & \prod_{i\geq 1} (1-t\,x_i)\,\,\,\,\,\,
= \, \sum_{m\geq 0}(-1)^m e_m(X) t^m\,,
\label{Eq-L}
\end{eqnarray}
where Schur functions $h_m(X)=s_{(m)}(X)$ and $e_m(X)=s_{(1^m)}(X)$.

\begin{remark}
There are some expansions which are differ from power series expansions 
that are useful in imperical studies {\rm(}for a detailed description see 
{\rm \cite{BCST,Andrews86})}. 
Indeed the following result holds 
\begin{eqnarray}
\prod_{n=\ell}^\infty(1-q^{an+\varepsilon})^{bn} & = &  {\mathcal R}
(s= (a\ell+\varepsilon)(1-i\varrho(\vartheta)) + 1-a)^{b\ell}
\nonumber \\
& \times &
\prod_{n=\ell+1}^\infty{\mathcal R}(s= (an+\varepsilon)(1-i\varrho(\vartheta)) + 1-a)^{b},
\end{eqnarray}
\begin{eqnarray} 
&& {\mathcal F}(X)^{a_n} = \prod_{n=1}^\infty (1-q^n)^{-a_n}  =  1+ \sum_{n=1}^\infty \mB_n q^n,
\label{prod}
\\
&& n\mB_n  =  \sum_{j=1}^nD_j\mB_{n-j} q^n, \,\,\,\,\, D_j = \sum_{d\vert j}da_d.
\label{prod1}
\end{eqnarray}
Here $a_n$ and $\mB_n$ are integers, function ${\mathcal R(s)}$ is an alternating product
of more complicate factors -- Patterson-Selberg spectral zeta-functions.
Note that if either sequance $a_n$ or $\mB_n$ is 
given, the other is uniquely determined by {\rm (\ref{prod1})}. 
\end{remark}

{\bf $Q$-symmetric functions.}
Let us introduce some more symmetric functions, which are called $Q$-functions.
The original definition for $Q_{(\lambda_1,\ldots,\lambda_p)}(x_1, \ldots, x_n)$
for a finite number of arguments is \cite{Schur}:
\begin{equation}
Q_{(\lambda_1,\ldots,\lambda_p)}(x_1, \ldots, x_n)\,
\stackrel{\rm def}{=} \, 2^p\!\! \sum_{j_i,\ldots,j_p =1}^n\frac{x_{j_1}^{\lambda_1}
\cdots x_{j_p}^{\lambda_p}}
{u_{j_1}\cdots u_{j_p}}{\mathcal A}(x_{j_p}, \ldots, x_{j_2}, x_{j_1}),
\end{equation}
where
\begin{equation}
{\mathcal A}(y_1, \ldots, y_p) = \prod_{1\leq i<j\leq p}\frac{y_i - y_j}{y_i + y_j},
\,\,\,\,\,\,\,
u_j = \prod_{1\leq i\leq n, i\neq j} \frac{x_j - x_i}{x_j + x_i}\,.
\end{equation}

{\bf The Hall-Littlewood functions.}
Note the generalization of the idea of symmetric functions, the Hall-Littlewood 
function \cite{Littlewood} in the variables $x_1, x_2, \ldots, x_n$ defined for 
a partition of length $\ell(\lambda)\leq n$:
\begin{equation}
Q_\lambda(x_1, \ldots, x_n; t)\,
\stackrel{\rm def}{=} \, (1-t)^{\ell(\lambda)}\!\! \sum_{\sigma\in S_n}\sigma
\left(x_1^{\lambda_1} \cdots x_n^{\lambda_n}\prod_{1\leq i<j\leq n}
\frac{x_i-tx_j}{x_i-x_j}\right)\,,
\label{HL}
\end{equation}
where $\sigma$ acts as $\sigma(x_1^{\lambda_1} \cdots x_n^{\lambda_n}) = 
x_{\sigma(1)}^{\lambda_1} \cdots x_{\sigma(n)}^{\lambda_n}$,
and $t$ is some parameter. When $t=0$, $Q_\lambda$ reduces to the $S$-function $s_\lambda$. 
Given a field $F$, let $\Lambda_F = \Lambda\otimes_{\mathbb Z}F$ be the ring of symmetric functions 
over F. In the case of the
Hall-Littlewood functions (\ref{HL}) in an infinite number of indeterminates, it is known 
\cite{Macdonald} that they form a basis for $F= {\mathbb Q}(t)$,
the field of rational functions in $t$. Functions $Q_\lambda(X)$ obey a Cauchy identity
\begin{equation}
\sum_{\lambda} 2^{-\ell(\lambda)}Q_\lambda(X)Q_\lambda(Y)  =
\prod_{i,j =1}^n \left(\frac{1+x_iy_j}{1-x_iy_j}\right)\,.
\end{equation}

{\bf The Jack symmetric functions.}
Another symmetric functions are the Jack symmetric functions $P_\lambda^{(\alpha)}(X)$, which are
defined as a particular limit of a Macdonald function
\begin{equation}
P_\lambda^{(\alpha)}(X) = \lim_{t\rightarrow 1}P_\lambda(X; t^\alpha,t)
\end{equation}
It has been poited out that the Jack symmetric functions $P_n^{(\alpha)}(X)$
can be expressed in the form $P_n^{(1/\alpha)}(X) =
(n!/\alpha^n)s_n(\alpha X)$, which specialize to zonal symmetric functions for $\alpha = 2$.
For these functions, there is an inner product $\langle \cdot , \cdot \rangle_\alpha$ on the
ring $\Lambda_G$ of symmetric functions with coefficients in $G = {\mathbb Q}(\alpha)$
which is defined by
\begin{equation}
\langle p_\lambda (X), p_\mu(X)) \rangle_\alpha =
\delta_{\lambda,\mu} \,\fz_\lambda \alpha^{\ell(\lambda)},
\label{ip}
\end{equation}
under which, the Jack symmetric functions obey the orthogonality relation
\begin{equation}
\langle P_\lambda^{(\alpha)}, P_\mu^{(\alpha)} \rangle_\alpha = \delta_{\lambda,\mu} \,j_\lambda,
\end{equation}
where calculation of the numerical factor $j_\lambda$ the reader can find in 
\cite{Stanley}, Theorem 5.8.
Let $g_n^{(\alpha)}(X) = P_{(n)}^{(\alpha)}(X)/j_{(n)}$ denote the elementary Jack function,
which has the generating function 
$
\sum_{n=0}^\infty g_n^{(\alpha)}(X)z^n = \prod_j (1-zx_j)^{1/\alpha}\,.
$
The Gram-Schmidt orthogonalization procedure gives a unique orthogonal basis for
$\Lambda_F$.

We shall be interested in the cases
\begin{equation}
\xi_n\, = \alpha\left(\frac{q^{\kappa n} - q^{-\kappa n}}{q^{2n} - q^{-2n}}\right) \, =  \,
\alpha [\kappa/2]_q = 
\alpha\left(\frac{\sin (2\pi\kappa\vartheta n)}{\sin (4\pi\vartheta n)}\right),
\label{DEF}
\end{equation}
where $\alpha \in {\mathbb R}$ and $\kappa\in {\mathbb Z}$. Recall that:

--\, The Hall-Littlewood symmetric functions correspond to the case when $\xi_n = (1-t^n)^{-1}$.

--\, The Maconald functions correspond to the case $\xi_n = (1-q^n)/(1-t^n)$.

Let us discuss set of symmetric functions over the field $F = {\mathbb Q}(q,t)$, which 
are generalizations of the Hall-Littlewood functions. Define an inner product on the power sum 
symmetric functions by
\begin{equation}
\langle p_\lambda(X), p_\mu(X)\rangle_{(q,t)} = 
\fz_\lambda(q,t)\delta_{\lambda,\mu},\,\,\,\,\,\,\,\,\,
\fz_\lambda(q,t) = 
\fz_\lambda\prod_{i=1}^{\ell(\lambda)}\frac{1-q^{\lambda_i}}{1-t^{\lambda_i}}\,.
\label{skew}
\end{equation}

Letting $b_\lambda^{-1}(q,t)\equiv \langle P_\lambda(q,t), P_\lambda(q,t)\rangle_{(q,t)}$,
define $Q_\lambda(q,t) = b_\lambda(q,t)P_\lambda(q,t)$ and use the following condition:
$\langle P_\lambda(X; q,t), P_\mu(X; q,t)\rangle_{(q,t)} = 0$, for $\lambda\neq \mu$. 
then we have
\begin{equation}
\langle P_\lambda(q,t), Q_\mu(q,t)\rangle_{(q,t)} = \delta_{\lambda,\mu}.
\label{PQ}
\end{equation}
Define
\begin{eqnarray}
\!\!\!\!\!\!\!\!\!\!\!\!\!\!
&&
(a; q)_n \stackrel{{\rm def}}{=} (1-a)(1-aq)\cdots (1-aq^{n-1});\,\,\,\,\,
(a; q)_\infty =\prod_{n=0}^\infty(1-aq^n);\,\,\,\, (a; q)_0 = 1,
\\
\!\!\!\!\!\!\!\!\!\!\!\!\!\!
&&
\Omega(tx_iy_j; \vartheta)  :=  {\rm log}(tx_iy_j)/ 2\pi i \vartheta\,.
\end{eqnarray}
From Eq. (\ref{PQ}), the functions $P_\lambda,\, Q_\lambda$ are dual basis for$\Lambda_F$;
it follows that the Macdonald functions $P_\lambda(X; q,t)$ obey the identity \cite{Macdonald}
\begin{eqnarray}\label{4.39}
\sum_\lambda \fz_\lambda^{-1}(q,t)p_\lambda(X)p_\lambda(Y) & = &
\sum_\lambda P_\lambda(X; q,t) Q_\lambda(Y; q,t)
\nonumber \\
& = &
\prod_{i,j}\frac{(tx_iy_j; q)_\infty}
{(x_iy_j; q)_\infty}\, = \prod_{i,j}\prod_{n=0}^\infty\frac{(1-tx_iy_jq^n)}{(1-x_iy_jq^n)}
\nonumber \\
& = &
\prod_{i,j}\prod_{n=0}^\infty\frac{(1-q^{n+\Omega(tx_iy_j; \vartheta)})}
{(1-q^{n+\Omega(x_iy_j; \vartheta)})}
\nonumber \\
& =&
\prod_{i,j}\frac{{\mathcal R}\left(s= \Omega(tx_iy_j; \vartheta)(1-i\varrho(\vartheta))\right)}
{{\mathcal R}\left(s= \Omega(x_iy_j; \vartheta)(1-i\varrho(\vartheta))\right)}\,,
\end{eqnarray}
with the last equality in \eqref{4.39} obtained by using the relation \eqref{R}.

Let us define the dual functions
$
Q_\lambda(X; q,\kappa,\alpha) \stackrel{{\rm def}}{=}
b_\lambda(q,\kappa, \alpha)P_\lambda(X; q,\kappa, \alpha),
$ \,
$
b_\lambda(q,\kappa, \alpha) = \vert\vert P_\lambda(X; q,\kappa, \alpha)\vert\vert^{-2},
$
such that
$
\langle P_\lambda(X; q,\kappa, \alpha), Q_\mu(X; q,\kappa, \alpha)\rangle = \delta_{\lambda\mu}.
$
From definition (\ref{DEF}) we see that:

-- $\lim_{q\rightarrow 1^{-}}P_\lambda(X; q, \kappa,\alpha) = P_\lambda^{(\kappa\alpha/2)}(X)$.

-- When $\kappa = 2$, the symmetric functions are Jack symmetric functions for all values of $q$.

-- When $\alpha = 1$, $P_\lambda(X; q,\kappa, \alpha)$ are identical to the Macdonald's function
$P_\lambda(q^{2\kappa}, q^4)$.

Let us now develop a Cauchy formula for the functions
$P_\lambda(X; q,\kappa, \alpha)$. Firstly, we have the result
\begin{eqnarray}
\!\!\!\!\!\!\!\!\!\!
\sum_\lambda \fz_\lambda^{-1}(q, \kappa,\alpha)p_\lambda(X)p_\lambda(Y) & = &
\exp\left(\frac{1}{\alpha}\sum_{n>0}\frac{\sin (2\pi\kappa\vartheta n)}{\sin (4\pi\vartheta n)}
p_n(X)p_n(Y)\right)
\nonumber \\
& = &
\prod_{i,j}\frac{(x_iy_jq^{\kappa+2}; q^{2\kappa})_\infty^{1/\alpha}}
{(x_iy_jq^{\kappa-2}; q^{2\kappa})_\infty^{1/\alpha}}
\nonumber \\
& \stackrel{by\,\,(\ref{R})}{=\!=\!=\!=\!=} &
\prod_{i,j}\left(\frac{{\mathcal R}(s= (\Omega(x_iy_j; \vartheta)+2)(1-i\varrho(\vartheta))-2)}
{{\mathcal R}(s= (\Omega(x_iy_j; \vartheta)-2)(1-i\varrho(\vartheta)))}\right)^{1/\alpha}
\!\!\!\!\!\!\!\!,
\label{standard}
\end{eqnarray}
where we have denoted $\fz_\lambda(q, \kappa,\alpha) = \fz_\lambda\xi_\lambda$ for the particular
choice (\ref{DEF}) of $\xi_\lambda$, and $(x;q)_\infty^a = \prod_{j=0}^\infty(1-xq^j)^a$.
Equation (\ref{standard}) is proved by a standard calculation (see \cite{Macdonald} 
for example). From this we obtain the following Cauchy identity
\begin{eqnarray}
\!\!\!\!\!\!\!\!\!\!\!\!\!\!\!
\sum_\lambda P_\lambda(X; q,\kappa, \alpha), Q_\lambda(X; q,\kappa, \alpha)\!\!\!\!\!\!\!\!\! & = &
\!\!\!\!\!\!\!\!\!
\prod_{i,j}\frac{(x_iy_jq^{\kappa+2}; q^{2\kappa})_\infty^{1/\alpha}}
{(x_iy_jq^{\kappa-2}; q^{2\kappa})_\infty^{1/\alpha}}
\nonumber \\
& \stackrel{by\,\,(\ref{R})}{=\!=\!=\!=\!=}  &
\prod_{i,j}\left(\frac{{\mathcal R}(s= (\Omega(x_iy_j; \vartheta)+2)(1-i\varrho(\vartheta))-2)}
{{\mathcal R}(s= (\Omega(x_iy_j; \vartheta)-2)(1-i\varrho(\vartheta)))}\right)^{1/\alpha}
\!\!\!\!\!\!\!\!.
\label{Cauchy}
\end{eqnarray}
The functions $P_\lambda(X; q,\kappa, \alpha)$ form a basis for
the ring $\Lambda_F$, so there exist structure constants
$f_{\mu\nu}^\lambda \equiv f_{\mu\nu}^\lambda(q\kappa,\alpha)$
(actually rational functions of the indeterminates $q$ and
$\alpha$) such that
\begin{eqnarray}
P_\mu(X; q,\kappa, \alpha)P_\lambda(X; q,\kappa, \alpha) & = & \sum_\lambda f_{\mu\nu}^\lambda
P_\lambda(X; q,\kappa, \alpha), \,\,\,\,\,\,\, {\rm or \,\,\,\, equivalently}
\\
Q_\mu(X; q,\kappa, \alpha)Q_\nu(X; q,\kappa, \alpha) & = &
\sum_\lambda {\overline f}_{\mu\nu}^\lambda Q_\lambda(X; q,\kappa, \alpha)\,,
\end{eqnarray}
where
$
{\overline f}_{\mu\nu}^\lambda = (b_\mu(q,\kappa,\alpha)b_\nu(q,\kappa,\alpha)/
(b_\lambda(q,\kappa,\alpha)) f_{\mu\nu}^\lambda\,.
$
It then follows from (\ref{Cauchy}) the folloing indeterminates
\begin{eqnarray}
P_\lambda(X,y; q,\kappa, \alpha) & = & \sum_\sigma P_{\lambda/\sigma}(X; q,\kappa, \alpha)
P_\sigma(X; q,\kappa, \alpha),
\\
Q_\lambda(X,y; q,\kappa, \alpha) & = & \sum_\sigma Q_{\lambda/\sigma}(X; q,\kappa, \alpha)
Q_\sigma(X; q,\kappa, \alpha).
\end{eqnarray}

Introduce the symmetric function of a replicated argument.
In order to define the function $P_\lambda(X^{(\tau)}; q,\kappa, \alpha)$, $\tau = m$, an integer, 
we put
\begin{equation}
P_\lambda(X^{(\tau)}; q,\kappa, \alpha) :=
P_\lambda(\overbrace{x_1, \ldots, x_1}^m,\overbrace{x_2,\ldots, x_2}^m, \ldots ; q,\kappa, \alpha).
\end{equation}
We introduce also the transition matrix
$Y_\lambda^\mu \equiv Y_\lambda^\mu(q,\kappa,\alpha)$ between the power sums and the functions 
$P_\lambda$,
\begin{equation}
p_\lambda(X) = \sum_\mu Y_\lambda^\mu P_\mu(X; q,\kappa, \alpha)\,.
\end{equation}
The functions $Y_\lambda^\mu$ have been studied in \cite{Srinivasan} in the case $\alpha = 1$
(the Macdonald case). From the Cauchy identities (\ref{standard}) and (\ref{Cauchy}), 
it follows that we have orthogonality relations of the form
\begin{eqnarray}
\sum_\rho \fz_\rho^{-1}(q,\kappa,\alpha) Y_\rho^\lambda Y_\rho^\mu & = & 
b_\lambda(q,\kappa, \alpha)
\delta_{\lambda\mu},
\label{1-ort}
\\
\sum_\lambda b_\rho^{-1}(q,\kappa,\alpha) Y_\rho^\lambda Y_\sigma^\mu & = & 
\fz_\rho(q,\kappa, \alpha)
\delta_{\rho\sigma}.
\label{2-ort}
\end{eqnarray}
$Y_\mu^{(n)} = 1$ for all partitions $\mu\vdash n$'n. The Cauchy identity is
\begin{eqnarray}
&&
\sum_\lambda P_\lambda(X^{(\tau)}; q, \kappa, \alpha)Q_\lambda(Y^{(\eta)}; q, \kappa, \alpha)
 =
\prod_{i, j}\left(\frac{(x_iy_jq^{\kappa +2}; q^{2\kappa})^{1/\alpha}_\infty}
{(x_i, y_j, q^{\kappa-2}; q^{2\kappa})^{1/\alpha}_\infty}\right)^{\tau\eta/\alpha}
\nonumber \\\\
\stackrel{by\,\,(\ref{R})}{=\!=\!=\!=\!=} 
\!\!\!\!\!\!
&&
\prod_{i,j}\left(\frac{{\mathcal R}(s= (\Omega(x_iy_j; 
\vartheta)+2)(1-i\varrho(\vartheta))-2)^{1/\alpha}}
{{\mathcal R}(s= (\Omega(x_iy_j; 
\vartheta)-2)(1-i\varrho(\vartheta))-2)^{1/\alpha}}\right)^{\tau\eta/\alpha}\!\!\!\!\!\!\!\!,
\end{eqnarray}
and therefore
\begin{eqnarray}
\sum_{n=0}^\infty Q_{(n)}(X; q, \kappa,\alpha)z^n & = &
\exp \left( \frac{1}{\alpha}\sum_{n>0} \frac{q^{\kappa n} - q^{-\kappa n}}
{q^{2n} - q^{-2n}}p_n(X)p_n(Y)\right)
\nonumber \\
& = &
\prod_i \left(\frac{(x_izq^{\kappa+2}; q^{2\kappa})_\infty}
{(x_izq^{\kappa-2}; q^{2\kappa})_\infty}\right)^{1/\alpha}
\nonumber \\
& \stackrel{by\,\,(\ref{R})}{=\!=\!=\!=\!=}  &
\prod_{i}\left(\frac{{\mathcal R}(s= (\Omega(x_iz; \vartheta)+2)(1-i\varrho(\vartheta))-2)}
{{\mathcal R}(s= (\Omega(x_iz; \vartheta)-2)(1-i\varrho(\vartheta))-2)}\right)^{1/\alpha}.
\end{eqnarray}

\subsection{Vertex operator traces}
\label{vertex}

Vertex operators have plays a fruitful role in string theory, quantum field theory, mathematical 
constructions of group representations as well as combinatorial constructions. We cite their 
applications to affine Lie algebras~\cite{Lepowsky,Frenkel80}, quantum affine algebras. 
Variations on the theme of symmetric functions are applications, for example, to $Q$-functions, 
Hall-Littlewood functions, Macdonald functions,
Jack functions (see in particular Sect. \ref{skew functions}), Kerov's symmetric functions  
(and a specialization of $S$-functions introduced by Kerov \cite{Kerov}).
By considering different specializations of Kerov's symmetric
functions the trace calculations in representations of the levels
quantum affine algebra $U_q({\mathfrak g}{\mathfrak l}_N)$ (see \cite{BC} for appropriate results)
can be feasible. The extension of this mathematical tools to other
(quantum) affine algebras and superalgebras is also practicable
and provides the relevant vertex operator realizations of those
algebras. 

Note that any irreducible highest weight representation of a Kac-Moody algebra can be
constructed as the quotient of a Verma module by its maximal proper submodule.
This construction suffices for some purposes, but in some cases other constructions
are known which give a connection with physics \cite{Feingold,Frenkel81}. In some cases this 
construction is known as the {\it vertex} \cite{Lepowsky,Frenkel80,Kac81}. Any Kac-Moody algebra has a root
system, a Weyl group, simple roots, and highest representations. In the affine case,
this gives the famous Macdonald identities for powers of the Dedekind $\eta$-function
\cite{Macdonald72}.

We consider here a general vertex operator which describe the currents of this realization, 
and which is able to connect with the symmetric and spectral functions. We specially note that 
realizations of (homogeneous) vertex operators are important in the high level representations 
theory of quantum affine algebras. Define generalized vertex operators as
\begin{eqnarray}
V(\overrightarrow{\tau}\ast Z; \,\overrightarrow{\eta}\ast W; \,\xi)
& = & \exp \left( \sum_{m>0} \frac{1}{m\xi_m}p_m(\tau_1z_1^m + \cdots + \tau_nz_n^m)\right)
\nonumber \\
& \times &
\exp \left( \sum_{m>0} \frac{1}{m\xi_m}D(p_m)(\eta_1 w_1^m + \cdots + \eta_n w_n^m)\right)\!,
\end{eqnarray}
where $D$ is the adjoint operator with respect to the inner product (\ref{skew}). That is,
$D(p_m) = m \xi_m \partial/\partial p_m$.

As an approach to generalising the vertex operators, the observations made in
previous sections allow us to write down expressions for replicated or parameterized vertex 
operators. In the simplest case, this is examplified by
\begin{eqnarray}
 V(\alpha z; -\alpha z^{-1}; 1) & := & V_\alpha(z) = {\mathcal F}(\alpha z;X)\, {\mathcal G}(\alpha z^{-1};X)
\nonumber \\
& = & \exp\left(  \alpha \sum_{k\geq1}\frac{z^k}{k}\, p_{k}\,\right)
\exp  \left(-\alpha\sum_{k\geq1}\, z^{-k}\, \frac{\partial}{\partial p_k}
\right) \,.
\label{simple}
\end{eqnarray}
for any $\alpha$, integer, rational, real or complex. 
In its simplest form (\ref{simple}) the vertex construction gives a representation 
$\hat{\mathfrak g}$ for $\mathfrak g$ of type $A, D$ or $E$ from the even integral 
root lattice $\Lambda$ of $\mathfrak g$ \cite{Cornwell}.
Then we have:
\begin{eqnarray}
 {\mathcal F}(\alpha z;X) & = &  {\mathcal F}(z;X)^\alpha = \prod_{i\geq1} (1-z\,x_i)^{-\alpha}
= \sum_{\sigma}\, s_\sigma(\alpha z)\, s_\sigma(X)
\nonumber \\
& = &  \sum_{\sigma}\, z^{|\sigma|} \dim_\sigma(\alpha)\ s_\sigma(X)\,,
\\
{\mathcal G}(\alpha z^{-1};X)
& = & {\mathcal G}(z^{-1};X)^\alpha = \prod_{i\geq1} (1-z^{-1}\,x_i)^\alpha
 = \sum_{\tau}\, (-1)^{|\tau|}\, s_{\tau}(\alpha z^{-1})\, s_{\tau'}(X)
   \nonumber \\
  & = & \sum_{\tau}\, (-z)^{-|\tau|} \dim_{\tau}(\alpha)\ s_{\tau'}(X)\,,
\end{eqnarray}
as given first in~\cite{Jarvis}. 
The Cauchy kernel ${\mathcal F}(XZ)$ serves as a generating function for characters of $GL(n)$ in the sense
that
\begin{equation}
{\mathcal F}(XY) = \prod_{i,j}(1-x_iy_j) = \sum_\lambda s_\lambda(X)s_\lambda(Y),
\end{equation}
where $s_\lambda(X)$ is the character of the irreducible representation $V_{GL(n)}^\lambda$
of highest weight $\lambda$ evaluated at group elements whose eigenvalues are the element of $X$.
We summarize some useful formulas:
\begin{eqnarray}
{\mathcal F}(q;XY) & = &  \prod_{i,j}(1-qx_iy_j)^{-1}  =  \sum_\alpha q^\alpha s_\alpha(X)s_\alpha(Y),
\nonumber \\
{\mathcal G}(q;XY) & = & \prod_{i,j}(1-qx_iy_j) \,\,\,\, = 
\,\,\, \sum_\alpha (-q)^{|\alpha|} s_\alpha(X)s_{\alpha^\prime}(Y).
\end{eqnarray}

Following the standard calculation \cite{Jarvis}, remark that the matrix
elements of the above vertex operator in a basis of Kerov symmetric functions take the form
\begin{equation}
\langle P_\mu (X; \xi), V Q_\nu (X; \xi)\rangle = \sum_\zeta P_{\mu/\zeta}
(\overrightarrow{\tau}\ast Z;\, \xi)\,
Q_{\nu/\zeta} (\overrightarrow{\eta}\ast W;\, \xi)\,.
\end{equation}
Suppose we want to calculate the regularized trace of the vertex operator $V$ over the
space $\Lambda_F$F. Define \cite{Jarvis}, 
\begin{eqnarray}
S_{p/r} & = & 
\sum_{\mu\nu} p^{|\mu|}r^{|\nu|} P_{\mu/\nu} 
(\overrightarrow{\tau}\ast Z;\, \xi)\, Q_{\mu/\nu} (\overrightarrow{\eta}\ast W;\, \xi)\,,
\\
A_{\lambda\mu} & = & \sum_{\zeta} p^{|\zeta|} P_{\zeta/\lambda}
(\overrightarrow{\tau}\ast Z;\, \xi)\, Q_{\zeta/\mu} (\overrightarrow{\eta}\ast W;\, \xi)\,.
\end{eqnarray}
Suppose that the Kerov functions with replicated arguments obey a very general
Cauchy identity
\begin{equation}
\sum_\lambda r^{|\lambda|} P_\lambda (X^{(\tau)}; \xi)\,\, Q_\lambda (Y^{(\eta)}; \xi)
= J_r^{\tau\eta}(X, Y; \xi)\,,
\end{equation}
so that for the functions $P_\lambda (X; \xi)$ with $\xi_\lambda$ defined by (\ref{DEF}) 
for example, the expression on the right has the form
\begin{eqnarray}
J_r^{\tau\eta}(X, Y; \xi) & = & \prod_{i,j}\left(\frac{(x_iy_jq^{\kappa+2}r; q^{2\kappa})_\infty}
{(x_iy_jq^{\kappa-2}r; q^{2\kappa})_\infty} \right)^{\tau\eta/\alpha}
\nonumber \\
& \stackrel{by\,\,(\ref{R})}{=\!=\!=\!=\!=} &
\prod_{i,j}\left(\frac{{\mathcal R}(s= (\Omega(x_iy_jr; \vartheta)+2)(1-i\varrho(\vartheta))-2)}
{{\mathcal R}(s= (\Omega(x_iy_jr; 
\vartheta)-2)(1-i\varrho(\vartheta))-2)}\right)^{\tau\eta/\alpha}\!\!\!\!.
\end{eqnarray}
We then form the generating function
${\mathcal J} = \sum_{\lambda\mu}A_{\lambda\mu}\,P_\lambda ({\mathfrak A})\,Q_\mu ({\mathfrak B})$, obtaining
\begin{eqnarray}
{\mathcal J} & = &   \sum_{\zeta} p^{|\zeta|} P_{\zeta}
(\overrightarrow{\tau}\ast Z,\,{\mathcal A};\, \xi)
\,\, Q_{\zeta} (\overrightarrow{\eta}\ast W,\, {\mathcal B};\, \xi)
\nonumber \\
& = &
\prod_{i,j = 1}^n J_p^{\tau_i\eta_j}(z_i, w_j; \xi)
\prod_{k = 1}^n J_p^{\tau_k,1}(z_k, {\mathfrak B}; 
\xi) J_p^{1, \eta_k}({\mathfrak A}, w_k; \xi)J_p^{1, 1}({\mathfrak A}, {\mathfrak B}; \xi)
\nonumber \\
& = &
\prod_{i,j = 1}^n J_p^{\tau_i\eta_j}(z_i, w_j; \xi)
\sum_{\sigma, \sigma_1,\sigma_2, \lambda, \mu}
p^{|\sigma|+ |\sigma_1|+ |\sigma_2|}P_{\sigma_1} (\overrightarrow{\tau}\ast Z;\, \xi)
\nonumber \\
& \times &
Q_{\sigma_2} (\overrightarrow{\eta}\ast W;\, \xi)f_{\sigma_2\sigma}^\lambda
P_\lambda ({\mathfrak A}; 
\xi) {\overline f}_{\sigma_1\sigma}^\mu Q_\mu ({\mathfrak B}; \xi)\,.
\end{eqnarray}
Finally we get (see also \cite{King90})
\begin{eqnarray}
A_{\lambda\mu} & = &  \prod_{i,j = 1}^n J_p^{\tau_i\eta_j}(z_i, w_j; \xi)
\sum_{\sigma}
p^{|\lambda|+ |\mu|- |\sigma|}P_{\mu/\sigma} (\overrightarrow{\tau}\ast Z;\, \xi)
\,\, Q_{\lambda/\sigma} (\overrightarrow{\eta}\ast W;\, \xi),
\\
S_{p/r} & = & \sum_\nu r^{|\nu|} A_{\nu\nu} = 
\prod_{i, j = 1}^n J_p^{\tau_i\eta_j}(z_i, w_j; \xi)
S_{rp^2/p^{-1}}\,.
\end{eqnarray}

In the Hall-Littlewood case, the above trace calculation leads to the particular
identities
\begin{eqnarray}
&&
\sum_{\mu\nu} q^{|\mu|} P_{\mu/\nu} (X^{(\alpha)}, Y^{(\beta)}; q)\,\,
Q_{\mu/\nu} (W^{(\tau)}, Z^{(\eta)}; q)  =
\prod_{n=1}^\infty (1-q^n)^{-1} \prod_{i, j}(1-qx_iw_j)^{-\alpha\tau}
\nonumber \\
&& \times \,\,
\prod_{k,l}(1-qx_kz_l)^{-\alpha\eta}
\prod_{m,n}(1-qy_mw_n)^{-\beta\tau}
\prod_{r,p}(1-qy_rz_s)^{-\beta\eta} \stackrel{by\,\,(\ref{R})}{=\!=\!=\!=\!=}  {\mathcal R}(s= 1-i\varrho(\vartheta))^{-1}
\nonumber \\
&& \times \,\,
{\mathcal F}(-\alpha\tau q;XW){\mathcal F}(-\alpha\eta q;XZ){\mathcal F}(-\beta\tau q;YW)
{\mathcal F}(-\beta\eta q;YZ)\,.
\end{eqnarray}


\section*{Asknowledgments}

We would like to thank Loriano Bonora for fruitful discussions on vertex operators and 
two-dimensional conformal field theory. AAB and RL would like to acknowledge the Conselho Nacional 
de Desenvolvimento Cient\'{i}fico e Tecnol\'{o}gico (CNPq, Brazil) and Coordenac\~{a}o de Aperfei\c{c}amento 
de Pessoal de N\'{i}vel Superior (CAPES, Brazil) for financial support.

\end{document}